\documentclass[prd,twocolumn,showpacs]{revtex4}
\usepackage{amsmath,amssymb,graphicx}

\begin{document}

\title{Stability of the Einstein static universe in $f(R)$ gravity}

\author{Christian G. B\"ohmer}
\email{christian.boehmer@port.ac.uk}
\affiliation{Institute of Cosmology \& Gravitation,
             University of Portsmouth, Portsmouth PO1 2EG, UK}

\author{Lukas Hollenstein}
\email{lukas.hollenstein@port.ac.uk}
\affiliation{Institute of Cosmology \& Gravitation,
             University of Portsmouth, Portsmouth PO1 2EG, UK}

\author{Francisco S.~N.~Lobo}
\email{francisco.lobo@port.ac.uk}
\affiliation{Institute of Cosmology \& Gravitation,
             University of Portsmouth, Portsmouth PO1 2EG, UK}

\date{\today}

\begin{abstract}

We analyze the stability of the Einstein static universe by
considering homogeneous scalar perturbations in the context of
$f(R)$ modified theories of gravity. By considering specific forms
of $f(R)$, the stability regions of the solutions are
parameterized by a linear equation of state parameter $w=p/\rho$.
Contrary to classical general relativity, it is found that in
$f(R)$ gravity a stable Einstein cosmos with a positive
cosmological constant does indeed exist. Thus, we are lead to
conclude that, in principle, modifications in $f(R)$ gravity
stabilizes solutions which are unstable in general relativity.

\end{abstract}

\maketitle

\section{Introduction}

Independent observations have confirmed that the Universe is
presently undergoing a phase of accelerated expansion
\cite{expansion}. Although the introduction of a cosmological
constant into the field equations seems to be the simplest
theoretical approach to generate a phase of accelerated expansion,
several alternative candidates have been proposed in the
literature, ranging from dark energy models to modified theories
of gravity. Amongst the latter, models generalizing the
Einstein-Hilbert action have been proposed. A nonlinear function
of the curvature scalar, $f(R)$, is introduced in the action given
by
\begin{equation}
S=\frac{1}{2\kappa^2}\int d^4x\sqrt{-g}\;f(R)+{\cal S}_m \,,
    \label{action}
\end{equation}
where ${\cal S}_m$ is the matter action. We consider $\kappa^2=
8\pi G=1$ throughout this work, for notational simplicity. Varying
the action with respect to $g_{\mu\nu}$ provides the following
field equation
\begin{equation}
FR_{\mu\nu}-\frac{1}{2}f\,g_{\mu\nu}-\nabla_\mu \nabla_\nu
F+g_{\mu\nu}\Box F=T^m_{\mu\nu} \,,
    \label{field:eq}
\end{equation}
where $F\equiv df/dR$. Note that the Ricci scalar is now a fully
dynamical degree of freedom, which is transparent from the
following relationship
\begin{equation}
FR-2f+3\,\Box F=T \,,
 \label{trace}
\end{equation}
obtained from the contraction of the modified field equation
(\ref{field:eq}). One may generalize the action (\ref{action}) by
considering an explicit coupling between an arbitrary function of
the scalar curvature, $R$, and the Lagrangian density of matter
\cite{Nojiri:2004bi}. Note that these couplings imply the
violation of the equivalence principle \cite{Olmo:2006zu}, which
is highly constrained by solar system tests. One may also mention
alternative approaches, namely, the Palatini formalism
\cite{Palatini,Sotiriou:2006qn}, where the metric and the
connections are treated as separate variables; and the
metric-affine formalism, where the matter part of the action now
depends on the connection and is varied with respect to it
\cite{Sotiriou:2006qn}.

A fundamental issue extensively addressed in the literature is the
viability of the proposed $f(R)$ models
\cite{viablemodels,Hu:2007nk}. In this context, it has been argued
that most $f(R)$ models proposed so far in the metric formalism
violate weak field solar system constraints \cite{solartests},
although viable models do exist
\cite{Hu:2007nk,solartests2,Sawicki:2007tf,Amendola:2007nt}
(Static and spherically symmetric solutions have also been found
\cite{SSSsol}). The issue of stability \cite{Faraoni:2006sy} also
plays an important role in the viability of cosmological solutions
\cite{Nojiri:2006ri,Sokolowski:2007pk,Amendola:2007nt}. In Ref.
\cite{Sawicki:2007tf} it was argued that the sign of
$f_{RR}=d^2f/dR^2$ determines whether the theory approaches the
general relativistic limit at high curvatures, and it was shown
that for $f_{RR}> 0$ the models are, in fact, stable. The
stability of the de Sitter solution in $f(R)$ gravity has also
been extensively analyzed in the literature \cite{deSitter}.
Recently, $f(R)$ models which have a viable cosmology were
analyzed, and it was found that the models satisfying cosmological
and local gravity constraints are practically indistinguishable
from the $\Lambda$CDM model, at least at the background level
\cite{Amendola:2007nt}. Note that to be a viable theory, the
proposed model, in addition, to simultaneously account for the
four distinct cosmological phases, namely, inflation, the
radiation-dominated and matter-dominated epochs, and the late-time
accelerated expansion \cite{Nojiri:2006be,Fay:2007gg}, should be
consistent with cosmological structure formation observations
\cite{structureform}. In the latter context, it has been argued
that the inclusion of inhomogeneities is necessary to distinguish
between dark energy models and modified theories of gravity, and
therefore, the evolution of density perturbations and the study of
perturbation theory  in $f(R)$ gravity is of considerable
importance \cite{Tsuj,Uddin:2007gj,Bazeia:2007jj}. See Ref.
\cite{growthfactor} for studies of a parameterized growth factor
approach to distinguish between modified gravity and dark energy
models using weak lensing.

In this work, we explore the stability of the Einstein static
universe in $f(R)$ modified theories of gravity. This is motivated
by the possibility that the universe might have started out in an
asymptotically Einstein static state, in the inflationary universe
context \cite{Ellis:2002we}. On the other hand, the Einstein
cosmos has always been of great interest in various gravitational
theories. In general relativity for instance, generalizations with
non-constant pressure have been analyzed in \cite{ESa}. In the
context of brane world models the Einstein static universe was
investigated in \cite{Gergely:2001tn}, while its generalization
within Einstein-Cartan theory can be found in
\cite{Boehmer:2003iv}. Finally, in the context of loop quantum
cosmology, we refer the reader to \cite{Mulryne:2005ef,foot1}.

We analyze the stability of the Einstein static universe against
homogeneous scalar perturbations in the context of $f(R)$ gravity.
In the following section, we provide two specific forms of $f(R)$,
and analyze the stability of the solutions, by considering
homogeneous scalar perturbations around the Einstein static universe.
The stability regions are given in terms of the linear equation of
state parameter $w=p/\rho$ and the unperturbed  energy density
$\rho_0$.

\section{The Einstein static Universe in $f(R)$ gravity}

Consider the metric given by
\begin{equation}
ds^2=-dt^2+a^2(t)\left[\frac{dr^2}{1-r^2}+r^2\,(d\theta^2 +\sin
^2{\theta} \, d\phi ^2)\right] \,.
    \label{Einst:metric2}
\end{equation}
For the Einstein static universe, $a=a_0=const$, the Ricci
scalar reduces to $R=6/a_0^2$, and the field equations take the
following form
\begin{eqnarray}
\rho_0=\frac{f}{2}\,, \qquad p_0=\frac{2F}{a_0^2}-\frac{f}{2}
\label{back:p0}\,,
\end{eqnarray}
where $\rho_0$ and $p_0$ are the unperturbed energy density and
isotropic pressure, respectively.

In what follows, we consider specific forms of $f(R)$, and analyze
the stability against linear homogeneous scalar perturbations
around the Einstein static universe given in Eqs.~(\ref{back:p0}).
Thus, we introduce perturbations in the energy density and the
metric scale factor which only depend on time:
\begin{equation}
  \rho(t) = \rho_0\left(1+\delta\rho(t)\right),
  \qquad
  a(t) = a_0\left(1+\delta a(t)\right).
  \label{def:perturbations}
\end{equation}
Subsequently, we consider a linear equation of state,
$p(t)=w\rho(t)$, linearize the perturbed field equations and
analyze the dynamics of the solutions.

Firstly, motivated by the possibility that the universe might have
started out in an asymptotically Einstein static state
\cite{Ellis:2002we}, we analyze the case of $f(R)\propto R+R^2$, as
in principle $R^2$ dominates for high curvatures, i.e.~the early
universe. Secondly, we consider the case of $f(R)\propto R+1/R$,
which is known to generate a late-time accelerated expansion phase
\cite{Carroll:2003wy}, and has been used in the weak-field limit
constraints, as the $1/R$ term dominates for low curvatures.

\vspace{.7cm}

\subsection{$f(R)\propto R+R^2$ theory}

Consider the case of
\begin{equation}
f(R)=R+\frac{\sigma\alpha^2 a_0^4}{6} R^2-2\Lambda\,,
\end{equation}
where $\sigma=\pm 1$, and $\alpha$ is a positive parameter. We
introduced the factor $a_0^4/6$ in the second term to considerably
simplify the equations and results in the analysis outlined below.

In this model the unperturbed field equations (\ref{back:p0}) take
the form
\begin{eqnarray}
\rho_0&=&\frac{3}{a_0^2}+3\sigma\alpha^2 -\Lambda
\,, \label{field4a} \\
p_0&=&-\frac{1}{a_0^2}+\sigma\alpha^2 +\Lambda  \,,
    \label{field4b}
\end{eqnarray}
and yield the following cosmological constant of the Einstein static
universe
\begin{equation}
\Lambda = \frac{1}{2}\rho_0(1+3w)-3\sigma\alpha^2 \,.
\end{equation}

We derive an evolution equation for the scale factor perturbation in
the following way. The perturbations defined in
Eq.~(\ref{def:perturbations}) are introduced in the metric and the
energy-momentum tensor. Then the perturbed field equations
(\ref{field:eq}) are linearized and the unperturbed equations
(\ref{field4a}) and (\ref{field4b}) are subtracted to end up having
only first order terms. The $(tt)-$component reduces to
\begin{equation}
\delta\rho(t)=-3(1+w)\delta a(t) \,,
    \label{tt}
\end{equation}
which we use to simplify the spatial component to
\begin{widetext}
\begin{multline}
\left[8\sigma\alpha^2-\rho_0(1+w)(1+3w)\right]
\left[-4\sigma\alpha^2+\rho_0(1+w)\right]^2\delta a(t) \\
-\,2\left[8\sigma\alpha^2-\rho_0(1+w)\right]
\left[4\sigma\alpha^2-\rho_0(1+w)\right] \delta a''(t)
+8\sigma\alpha^2 \delta a^{(4)}(t)=0 \,.
\label{perturbR2}
\end{multline}
\end{widetext}

Note that in the general relativistic limit, $\alpha\rightarrow
0$, Eq.~(\ref{perturbR2}) reduces to
\begin{equation}
2\delta a''(t)-\rho_0(1+w)(1+3w) \delta a(t)=0 \,,
\label{eq:grlimit}
\end{equation}
which provides the solution
\begin{equation}
\delta a(t)=C_1e^{\bar{\omega} t}+C_2e^{-\bar{\omega} t} \,,
\end{equation}
where $C_1$ and $C_2$ are constants of integration, and
$\bar{\omega}$ is defined by
\begin{equation}
\bar{\omega}=\sqrt{\frac{1}{2}\rho_0(1+w)(1+3w)} \,.
\end{equation}
In order to avoid the blow-up, due to the exponential increase in the
scale factor, or the collapse, the solution is stable in the range
\begin{equation}
-1<w<-1/3 \,.
\end{equation}
Note that this interval violates the strong energy condition which
stipulates that $\rho+3p \geq 0$.

Since the cosmological constant of the classical
Einstein universe is given by
\begin{equation}
\Lambda = \frac{1}{2}\rho_0(1+3w) \,,
\end{equation}
and we are only considering positive energy densities, it turns
out that it is {\it negative} in the region of stability.

Now, the full modified perturbation differential equation,
Eq.~(\ref{perturbR2}), provides the following solution
\begin{equation}
\delta a(t)=C_1e^{\omega_1 t}+C_2e^{-\omega_1 t}+C_3e^{\omega_2
t}+C_4e^{-\omega_2 t} \,,
\end{equation}
where $C_i$ (with $i=1...4$) are constants, and the parameters
$\omega_1$ and $\omega_2$ are given by
\begin{eqnarray}
&&\omega_{1,2}=\Big\{ \frac{\rho_0(1+w)-4\sigma\alpha^2}{8\sigma\alpha^2}
                       \Big[8\sigma\alpha^2-\rho_0(1+w) \pm
    \nonumber   \\
&&\hspace{-0.5cm}
\pm\sqrt{\rho_0(1+w)\left[\rho_0(1+w)+8\sigma\alpha^2(3w-1)\right]}\Big]\Big\}^{1/2}\,,
    \label{perturb3}
\end{eqnarray}
respectively.

In the following, we require the cosmological constant to be
positive. For $\sigma=-1$ and $\Lambda>0$ no stable solutions can be
found. Considering $\sigma=+1$ and $\Lambda >0$, we have the
following three stability regions: Firstly,
\begin{eqnarray}
\mbox{AI}:\qquad
8\alpha^2<&\rho_0&<\frac{3(7+\sqrt{17})\alpha^2}{2}
\,,         \\
\frac{8\alpha^2-\rho_0}{24\alpha^2+\rho_0}\leq
&w&<\frac{1}{3}\left(-2+\sqrt{\frac{24\alpha^2+\rho_0}
{\rho_0}}\right) \,.
    \label{stability1b}
\end{eqnarray}

Secondly,
\begin{eqnarray}
\mbox{AII}:\qquad \rho_0&=&\frac{3(7+\sqrt{17})\alpha^2}{2} \,,
         \\
-\frac{5+3\sqrt{17}}{3(23+\sqrt{17})}<
&w&<\frac{1}{3}\left(-2+\sqrt{\frac{23+\sqrt{17}}{7+\sqrt{17}}}\right)
\,.
    \label{stability2b}
\end{eqnarray}
The latter inequality reduces to $-0.213<w<-0.146$.

\begin{figure*}
\centering
\includegraphics[width=4.3in]{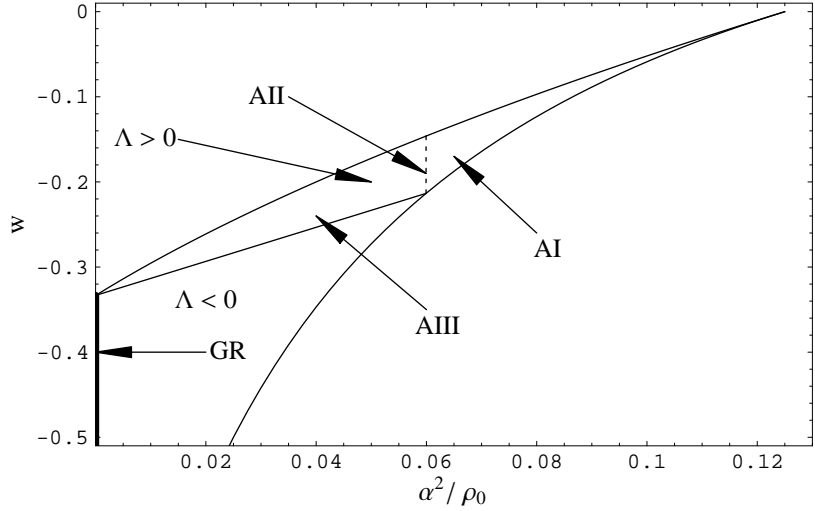}
\caption{The stability regions, for the case of $f(R)\propto
R+R^2$ theory, are depicted between the curves. The thick solid
line represents general relativity. In the lower left region,
which tends to general relativity, the cosmological constant is
negative. However, in the upper triangular-like shaped region (AI,
AII \& AIII) the Einstein static universe is stable and $\Lambda$
is positive, compare also with Fig.~\ref{Fig:plot2}. Note that the
equation of state parameter $w$ is strictly negative.}
\label{Fig:plot1}
\end{figure*}

Finally,
\begin{eqnarray}
\mbox{AIII}:\qquad \rho_0&>&\frac{3(7+\sqrt{17})\alpha^2}{2} \,,
         \\
\frac{6\alpha^2-\rho_0}{3\rho_0}\leq
&w&<\frac{1}{3}\left(-2+\sqrt{\frac{24\alpha^2+\rho_0}
{\rho_0}}\right) \,.
    \label{stability3b}
\end{eqnarray}
These stability regions are summarized in Table \ref{t1}, and
depicted in Fig.~\ref{Fig:plot1}. Note that these results are
consistent with the stability condition $f_{RR}=d^2f/dR^2>0$ of
cosmological models at high curvatures \cite{Sawicki:2007tf}.

\begin{table*}
\begin{center}
\begin{tabular}[c]{|c|c|c|}
\hline\hline Case AI &
\hspace{0.5cm}$8\alpha^2<\rho_0<\frac{3(7+\sqrt{17})\alpha^2}{2}$
\hspace{0.5cm}&
\hspace{0.5cm}$\frac{8\alpha^2-\rho_0}{24\alpha^2+\rho_0}\leq
w<\frac{1}{3}\left(-2+\sqrt{\frac{24\alpha^2+\rho_0}
{\rho_0}}\right)$ \hspace{0.5cm}  \\
\hline\hline
     Case AII & $\rho_0=\frac{3(7+\sqrt{17})\alpha^2}{2}$ & $-\frac{5+3\sqrt{17}}{3(23+\sqrt{17})}<
w<\frac{1}{3}\left(-2+\sqrt{\frac{23+\sqrt{17}}{7+\sqrt{17}}}\right)$ \\
\hline\hline Case AIII & $\rho_0>\frac{3(7+\sqrt{17})\alpha^2}{2}$ &
$\frac{6\alpha^2-\rho_0}{3\rho_0}\leq
w<\frac{1}{3}\left(-2+\sqrt{\frac{24\alpha^2+\rho_0}
{\rho_0}}\right)$   \\
 \hline\hline
\end{tabular}
\caption{Summary of the stability regions in the Einstein static
universe for the specific case of $f(R)\propto R+R^2$
theory.}\label{t1}
\end{center}
\end{table*}

\subsection{$f(R)\propto R+1/R$ theory}

In this section, we use the specific form of
\begin{equation}
f(R)=R+\frac{\sigma\mu^4}{a_0^2} \frac{1}{R}-2\Lambda\,,
   \label{f(R):1/R}
\end{equation}
where $\sigma=\pm 1$, and $\mu$ is considered a positive parameter.
As in the previous example, we have inserted the factor $a_0^2$ which
simplifies the calculations and the respective notation outlined
below.

The choice of the $f(R)$ given by Eq.~(\ref{f(R):1/R}) has been
extensively analyzed in the literature since it was first
demonstrated to account for the late-time accelerated expansion of
the Universe without the need for the introduction of dark energy
\cite{Carroll:2003wy}. It has also been used in the weak-field
limit, as now the $1/R$ term dominates for low curvatures.
Unfortunately, it was demonstrated that the originally proposed
form suffers from instabilities \cite{Dolgov:2003px}. However, it
was recently shown in \cite{Sawicki:2007tf} that a modification of
the sign stabilizes the solution such that $f_{RR}=d^2f/dR^2>0$,
as emphasized in the Introduction.

Considering this case, the unperturbed field equations (\ref{back:p0})
take the form
\begin{eqnarray}
\rho_0&=&\frac{3}{a_0^2}+\frac{\sigma\mu^4}{12} -\Lambda\,, \\
p_0&=&\frac{1}{a_0^2}-\frac{5\sigma\mu^4}{36}+\Lambda\,,
\label{field2}
\end{eqnarray}
which implies that the cosmological constant of the modified Einstein
static universe is given by
\begin{equation}
\Lambda = \frac{1}{2} \rho_0 (1+3w) + \frac{\sigma\mu^4}{6}\,.
\end{equation}

Applying linear perturbation theory, in the procedure outlined in
the previous case, we deduce the following differential equation
\begin{widetext}
\begin{multline}
\left[2\sigma \mu^4 -9(1+w)(1+3w)\rho_0\right]
\left[\sigma \mu^4 +18(1+w) \rho_0\right]^2 \delta a(t) \\
-36\left[2 \sigma \mu^4 -9(1+w)\rho_0\right]
\left[\sigma \mu^4 +18(1+w)\rho_0\right]
\delta a''(t) + 648 \sigma \mu^4 \delta a^{(4)}(t)=0 \,.
\label{perturb}
\end{multline}
\end{widetext}
In the limit $\mu \rightarrow 0$ this differential equations also
reduces to the general relativistic one, i.e.,
Eq.~(\ref{eq:grlimit}). Equation~(\ref{perturb}) provides the
following solution
\begin{equation}
\delta a(t)=C_1e^{\omega_3 t}+C_2e^{-\omega_3 t}+C_3e^{\omega_4
t}+C_4e^{-\omega_4 t} \,.
\end{equation}
The parameters $\omega_3$ and $\omega_4$ are given by
\begin{eqnarray}
&&\omega_{3,4}=\Big\{\frac{\mu^4\sigma+18\rho_0(1+w)}
{36\mu^4\sigma}\Big\{2\mu^4\sigma- 3\Big[3\rho_0(1+w)\pm
    \nonumber   \\
&&\hspace{-0.5cm}\pm\sqrt{\rho_0(1+w)\left[2\mu^4\sigma(-1+3w)
+9\rho_0(1+w)\right]}\Big]\Big\} \Big\}^{1/2}\,.
    \label{perturb2}
\end{eqnarray}

As in the previous example, we verify that for $\sigma=-1$ and a
positive cosmological constant, the solutions are unstable.
Considering $\sigma=+1$ and $\Lambda>0$, we have the following
stability regions
\begin{eqnarray}
\mbox{BI}:\qquad \frac{2\mu^4}{9}<&\rho_0&<\frac{(5+\sqrt{41})\mu^4}{12}\,, \\
\frac{2\mu^4-9\rho_0}{3(2\mu^4+3\rho_0)}\leq
&w&<\frac{1}{9}\left[-6+\sqrt{\frac{3(2\mu^4
+3\rho_0)}{\rho_0}}\right]   \,,.
    \label{stability}
\end{eqnarray}

Secondly,
\begin{eqnarray}
\mbox{BII}:\qquad \rho_0&=&\frac{(5+\sqrt{41})\mu^4}{12} \,, \\
-\frac{7+3\sqrt{41}}{3(13+\sqrt{41})}<
&w&<\frac{1}{3}\left(-2+\sqrt{\frac{13+\sqrt{41}}{5+\sqrt{41}}}
\right) \,.
    \label{stability2}
\end{eqnarray}

\begin{figure*}
\centering
\includegraphics[width=4.3in]{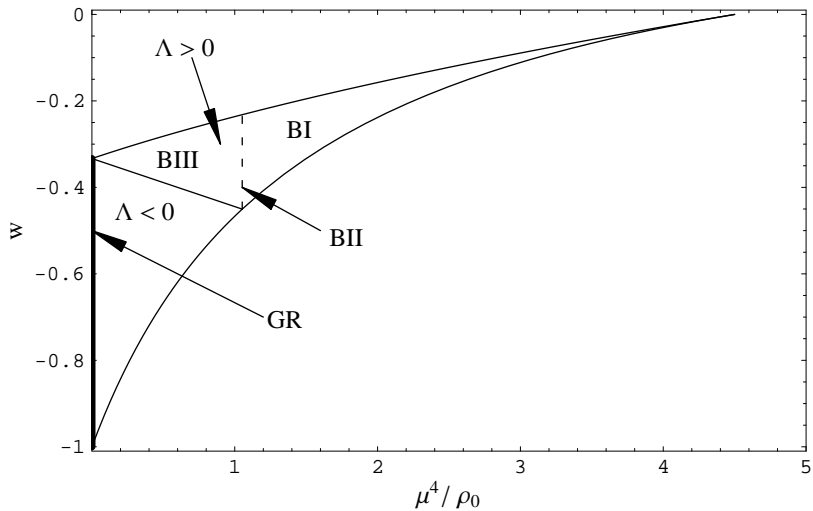}
\caption{The stability regions, for the case of $f(R)\propto R+1/R$
theory, are depicted between the curves. As before, the thick solid
line represents general relativity. In the lower left region, which
tends to general relativity, the cosmological constant is negative.
In the upper triangular-like shaped region (BI, BII \& BIII) the
Einstein static universe is stable and $\Lambda$ is positive. As
before, $w$ is strictly negative.}
\label{Fig:plot2}
\end{figure*}

Finally
\begin{eqnarray}
\mbox{BIII}:\qquad \rho_0&>&\frac{(5+\sqrt{41})\mu^4}{12} \,,
         \\
-\frac{\mu^4+3\rho_0}{9\rho_0}<
&w&<\frac{1}{9}\left[-6+\sqrt{\frac{3(2\mu^4
+3\rho_0)}{\rho_0}}\right] \,.
    \label{stability3}
\end{eqnarray}
We summarize these stability regions in Table \ref{t2}, and depict
the solutions in Fig.~\ref{Fig:plot2}. Note that, as emphasized
above, these results are also consistent with the stability
condition $f_{RR}=d^2f/dR^2>0$ \cite{Sawicki:2007tf}.
\begin{table*}
\begin{center}
\begin{tabular}[c]{|c|c|c|}
\hline\hline Case BI &
\hspace{0.5cm}$\frac{2\mu^4}{9}<\rho_0<\frac{(5+\sqrt{41})\mu^4}{12}$\hspace{0.5cm}
&\hspace{0.5cm}
$\frac{2\mu^4-9\rho_0}{3(2\mu^4+3\rho_0)}\leq
w<\frac{1}{9}\left[-6+\sqrt{\frac{3(2\mu^4
+3\rho_0)}{\rho_0}}\right] \hspace{0.5cm} $   \\
\hline\hline
     Case BII & $\rho_0=\frac{(5+\sqrt{41})\mu^4}{12}$ & $-\frac{7+3\sqrt{41}}{3(13+\sqrt{41})}<
w<\frac{1}{3}\left(-2+\sqrt{\frac{13+\sqrt{41}}{5+\sqrt{41}}}
\right)$ \\
\hline\hline Case BIII &
$\rho_0>\frac{(5+\sqrt{41})\mu^4}{12}$ &
$-\frac{\mu^4+3\rho_0}{9\rho_0}<
w<\frac{1}{9}\left[-6+\sqrt{\frac{3(2\mu^4
+3\rho_0)}{\rho_0}}\right]$   \\
 \hline\hline
\end{tabular}
\caption{Summary of the stability regions in the Einstein static
universe for the specific case of $f(R)\propto R+1/R$
theory.}\label{t2}
\end{center}
\end{table*}

\section{Summary and discussion}\label{sec:conclusion}

The Einstein static universe has recently been revived as the
asymptotic origin of an emergent universe, namely, as an
inflationary cosmology without a singularity \cite{Ellis:2002we}.
The role of positive curvature, negligible at late times, is
crucial in the early universe, as it allows these cosmologies to
inflate and later reheat to a hot big-bang epoch. An attractive
feature of these cosmological models is the absence of a
singularity, of an `initial time', of the horizon problem, and the
quantum regime can even be avoided. Furthermore, the Einstein
static universe was found to be neutrally stable against
inhomogeneous linear vector and tensor perturbations, and against
scalar density perturbations provided that the speed of sound
satisfies $c_{\rm s}^2>1/5$ \cite{Barrow:2003ni}. Further issues
related to the stability of the Einstein static universe may be
found in Ref. \cite{Barrow}

In this work we have analyzed linear homogeneous scalar
perturbations around the Einstein static universe in the context
of $f(R)$ modified theories of gravity. We have considered two
specific forms of $f(R)$ and found the stability regions of the
solutions for the scale factor perturbation. The first case
considered, namely, $f(R)\propto R+R^2$, was motivated by the fact
that, in principle, $R^2$ dominates for high curvatures which is
expected in the early universe. Secondly, we considered the case
of $f(R)\propto R+1/R$, which is known to generate a late-time
accelerated expansion phase \cite{Carroll:2003wy}, and has been
used in the weak-field limit, as now the $1/R$ term dominates for
low curvatures. The stability regions were parameterized by an
equation of state parameter $w=p/\rho$, and it was found that in the
context of $f(R)$ modified theories of gravity the range of the
parameter is greatly enhanced relatively to the results obtained in
general relativity. However, in both cases we analyzed, the equation
of state parameter was strictly negative, an issue which we hope to
overcome in future work by considering other modified gravity models.
These results are consistent with the condition $f_{RR}=d^2f/dR^2>0$,
for the stability of cosmological models \cite{Sawicki:2007tf}.

Concluding, we have found that the modified Einstein static universe,
with a positive cosmological constant and matter described by the
equation of state, $p=w\rho$, can be stabilized against homogeneous
perturbations, contrary to classical general relativity. Therefore,
we are lead to conclude that, in principle, stable modified gravity
solutions, which are unstable in general relativity \cite{foot2}, do
indeed exist. That this is actually possible relies on the fact that
the perturbations of the metric couple to the matter perturbations,
see our Eq.~(\ref{tt}). Similar results have been obtained in
\cite{foot1} where the modified Friedman equations in loop quantum
cosmology are considered.

\acknowledgments
The authors thank Roy Maartens, Luca Parisi and Sanjeev Seahra for
helpful discussions. The work of CGB was supported by research grant
BO 2530/1-1 of the German Research Foundation (DFG). FSNL was funded
by Funda\c{c}\~{a}o para a Ci\^{e}ncia e a Tecnologia (FCT)--Portugal
through the grant SFRH/BPD/26269/2006.

\end{document}